# FLUCTUATING INTERFACES, SURFACE TENSION, AND CAPILLARY WAVES: AN INTRODUCTION

VLADIMIR PRIVMAN

*Department of Theoretical Physics, University of Oxford*\*
*1 Keble Road, Oxford OX1 3NP, UK*

ABSTRACT
We present an introduction to modern theories of interfacial fluctuations and the associated interfacial parameters: surface tension and surface stiffness, as well as their interpretation within the capillary wave model. Transfer matrix spectrum properties due to fluctuation of an interface in a long-cylinder geometry are reviewed. The roughening transition and properties of rigid interfaces below the roughening temperature in $3d$ lattice models are surveyed with emphasis on differences in fluctuations and transfer matrix spectral properties of rigid vs. rough interfaces.

## 1. Interfaces at Phase Coexistence

This article presents a review of various concepts and ideas in theories of interfaces and their fluctuations. The presentation aims at providing a tutorial introduction emphasizing notation, nomenclature and uniformity of exposition of results accumulated in the field of Statistical Physics, on static interfacial fluctuation properties. Thus, no completeness of material is attempted, nor is the list of literature complete or even balanced. Instead, general citation of review articles for further reference is provided (Refs. 1-12).

Study of interfaces is an old topic dating back to first advances in Condensed Matter Physics. However, it grew immensely in scope and subject coverage recently, due to modern theoretical threatments with focus on fluctuation and phase-transition phenomena. Topics such as capillary waves, finite-size effects, wetting and surface phase transitions, roughening, connections with Polymer Physics and studies of membranes, etc., have each become a field of active research. The scope of the present review is limited to an overall introduction to selected topics, as well as summary of some recent results on the relations between interfacial fluctuations and transfer matrix spectral properties. As part of the conference Proceedings, and in fact, being a summary of a review talk presented, this survey avoids closely related topics covered in other presentations, notably, some finite-size effects. Instead, emphasis was put on issues of relevance in numerical studies of interfacial properties.

---

\*on leave of absence from Department of Physics, Clarkson University, Potsdam, NY 13699, USA

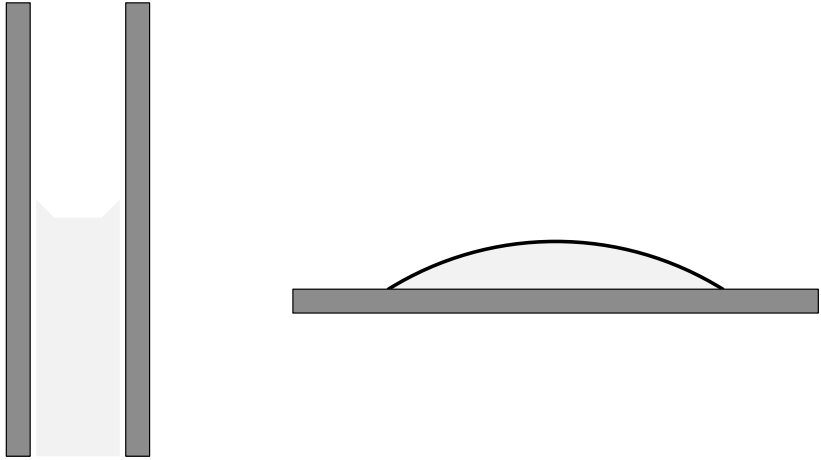

Fig. 1. Interface in a narrow capillary and bounding a liquid drop.

Static interfaces, or sharp domain walls, separate phases at coexistence. One has to assume that no gradual, continuous variation from one phase to another is possible, via a diffuse domain wall such as found, for instance, in $O(n)$ vector ferromagnets with continuous symmetry, $n = 2, 3, \ldots$, and the associated soft-mode spin-wave excitations. Thus, interfaces separate distinct phases such as $\pm$ magnetized ($n = 1$) Ising-model phases. In fact, walls and surfaces are all around us, and in the "zeroth-order approximation" that's all we actually see! However, *thermally equilibrated* (on experimental time scales) interfaces of the type Statistical Physics treats are less abundant. They can be observed between fluid phases, such as the gas-liquid meniscus in a narrow capillary, or surface of liquid drop on a wall; see Figure 1. The longest wave-length fluctuations are, however, typically cut off by gravity. True equilibrium crystal shapes can be also experimentally realized in some instances.

Theoretically, the issue of how to define interfaces microscopically, and how to measure their properties, has encountered difficulties related to the fact that interfacial fluctuation are rough, soft-mode, in many cases; details will be given later. Rough interface can be viewed as a sheet-like object, structureless and fluctuating, when it is "resolved" on a length scale large compared to the microscopic correlation lengths of the coexisting phases. This is illustrated by the broken-line box in Figure 2. A more microscopic resolution would reveal a region structured differently from each of the two phases in contact, and forming a boundary between the phases; see the solid-line box in Figure 2. Theoretical description of the structure in this boundary region in terms of an "intrinsic" interfacial profile, e.g., density profile for a liquid-gas interface, is intimately connected with such box-resolution, so-called coarse-graining notions, on various length scales.

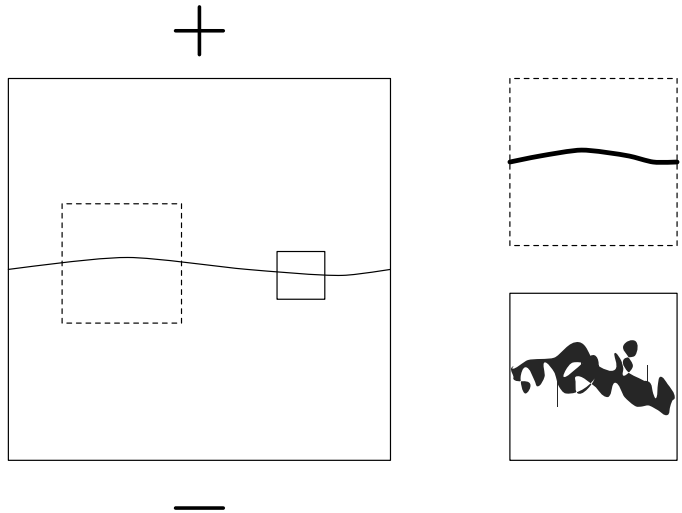

Fig. 2. Coarse-graining to reveal large-scale interfacial structure (broken-line box) and "intrinsic" microscopic structure (solid-line box), shown schematically.

The concept of "intrinsic" vs. "large-scale" interfacial structure has never been well defined outside the framework of a given theory. There is no clear "snapshot" definition of an interface. By examining a fixed system configuration, such as counting $+-$ Ising bonds, identifying $\pm$ clusters, etc., one cannot define an interface unambiguously. There will always be overhangs, islands, clusters, which may be assigned to the phase-separating interface region, or equally well attributed to the single-phase fluctuation structure surrounding the interface.

Thus, in order to define interface location and measure its properties, one appeals to external, essentially boundary-condition-based methods of fixing the interface. For instance, the interface in Figure 2 was secured by the $\pm$ boundary conditions at the upper and lower faces of the finite-size sample. If the sides are subject to periodic or free boundary conditions, then this interface will "float," oriented horizontally on the average. However, one can also pin the interfacial circumference by more restrictive boundary conditions. In Figure 3, the interface is forced to have average inclination angle $\theta$ with respect to the horizontal axis of the sample, aligned with the axes of the underlying lattice (drawn schematically in the lower right corner). There are instances of interfaces generated spontaneously due to energy gain (such as in microemulsions) or entropy gain (for instance, in long-cylinder geometries with periodic or free boundary conditions in the transverse directions). However, we restrict our attention to interfaces forced in by boundary conditions.

In a typical mixed-boundary-condition geometry there is an excess free-energy associated with the interface. For instance, in the geometries shown in Figures 2 and 3, the total free energy will be larger than in similar reference systems obtained by replacing all the $-$ boundary conditions by $+$. The total free-energy increase will be proportional to the interfacial area. Let $\Sigma$ denote the excess free energy per unit interfacial area, while $F$ refer to the total free energy of the finite-size sample. We define

$$\Sigma = \frac{F_{+-} - F_{++}}{L^{d-1}/\cos\theta} , \qquad (1.1)$$

where $L^{d-1}/\cos\theta$ is the interfacial area, with $\theta = 0$ for the case of Figure 2. The subscripts of the free energies indicate the boundary conditions, as just described.

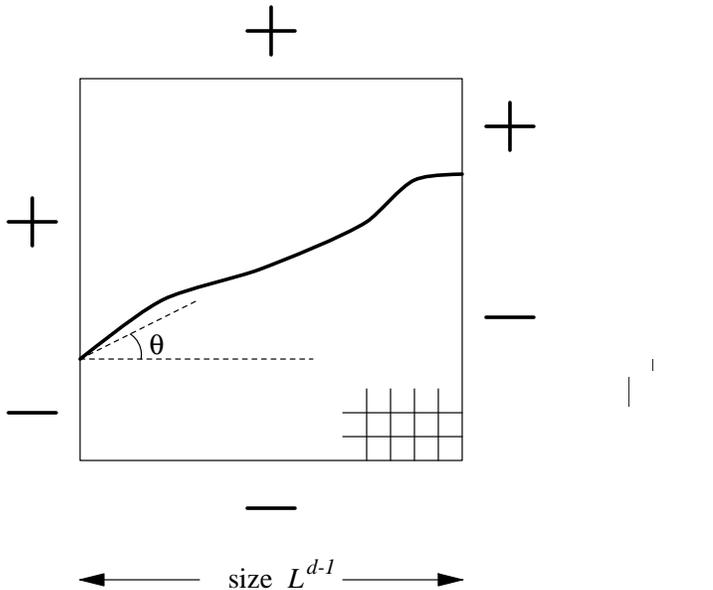

Fig. 3. Interface with ends pinned by boundary conditions, in a $2d$ sample. Or cross-section of pinned-circumference interface of average area $L^{d-1}/\cos\theta$ in a $d$-dimensional sample.

The interfacial free energy density, $\Sigma(T, \theta, L, \ldots)$, depends on system sizes, on shape ratios and boundary conditions. Most of these dependencies are, however, corrections typically of order $1/L$ or smaller. The leading, order 1, contribution in the thermodynamic limit of infinite sample, gives the macroscopic *surface tension*. Finite-size corrections in some geometries are surveyed in other contributions to these Proceedings.

It is convenient to introduce the reduced quantity, of units (area)$^{-1}$,

$$\sigma \equiv \Sigma/kT . \qquad (1.2)$$

An important observation is that interfaces usually separate phases which are non-critical and therefore microscopic fluctuations are not necessarily isotropic. Thus, not only the temperature dependence but also the *angle dependence* of the surface tension survives in the thermodynamic limit. The angular dependence is with respect to the underlying lattice structure, so that it is not present for fluids but plays an important role when at least one of the coexisting phases is solid. For model lattice systems used in numerical simulations, the finite-size samples are usually oriented along a symmetric lattice axis, as marked schematically in Figure 3. In $d$ dimensions, $(d-1)$ angular variables are needed to specify the angular dependence of $\sigma$.

## 2. Surface Tension and Surface Stiffness Coefficient

In this section we survey properties of the reduced bulk (i.e., infinite-system limiting) surface tension, $\sigma(T, \theta)$. We use the $2d$ notation with one angle, and whenever appropriate, quote results in the form relevant for rough interfaces. Both the concept of rough interfaces and the transition to the rigid-interface behavior at low temperatures in lattice $3d$ models will be further explained in later sections.

Interfaces are termed "rough" when due to gain in entropy they are not pinned by the underlying lattice structure. Their fluctuations are long-ranged, soft-mode. They can therefore pass locally at any angle with respect to the lattice axes. The angular dependence is smooth, analytic,

$$\sigma(\theta) = \sigma(0) + \sigma'(0)\theta + \frac{1}{2}\sigma''(0)\theta^2 + \dots . \tag{2.1}$$

As mentioned, in numerical studies the angle is always measured with respect to a symmetric axis, so that the term linear in $\theta$ vanishes.

In the geometry of Figure 3, let us consider the interfacial free energy per unit *projected*, horizontal area, $L^{d-1}$, cf. (1.1),

$$\sigma(\theta)/\cos\theta = \sigma(0) + \sigma'(0)\theta + \frac{1}{2}\kappa\theta^2 + \dots , \tag{2.2}$$

where we defined an important quantity

$$\kappa \equiv \sigma + \frac{d^2\sigma}{d\theta^2} , \tag{2.3}$$

termed the surface stiffness coefficient. This nomenclature will become clear in the next section, when we describe the capillary wave model.

In two dimensions, interfaces are always rough, for $0 < T < T_c$, where we use the Ising-model notation for the $\pm$ phase coexistence in zero applied field below the Curie temperature $T_c$. As shown in Figure 4, the $2d$ surface tension vanishes according to

$$\sigma \sim (T_c - T)^\mu , \tag{2.4}$$

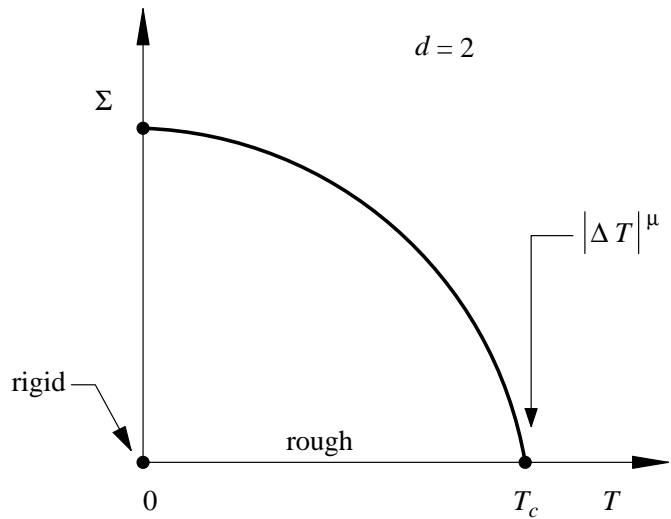

Fig. 4. Surface tension of $d = 2$ systems.

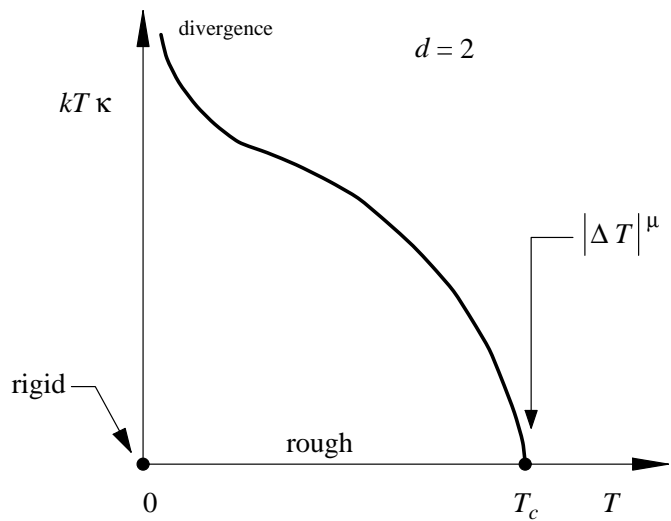

Fig. 5. Surface stiffness coefficient of $d = 2$ systems.

as $T \to T_c^-$, with the exponent values typically $\mu = 1$ or smaller (infinite slope), depending on a system universality class. At $T = 0$, the interfaces are rigid even in $2d$. This concept will be described in detail in Section 5 below. For now it suffices to note that rigid interfaces do follow lattice-axis directions microscopically, pinned by energy constraints at low temperatures when entropy gain plays no important role. The energy cost of the interface is finite at low temperatures; the unreduced quantity $kT\sigma = \Sigma$, shown in Figure 4, approaches a finite value at $T = 0$.

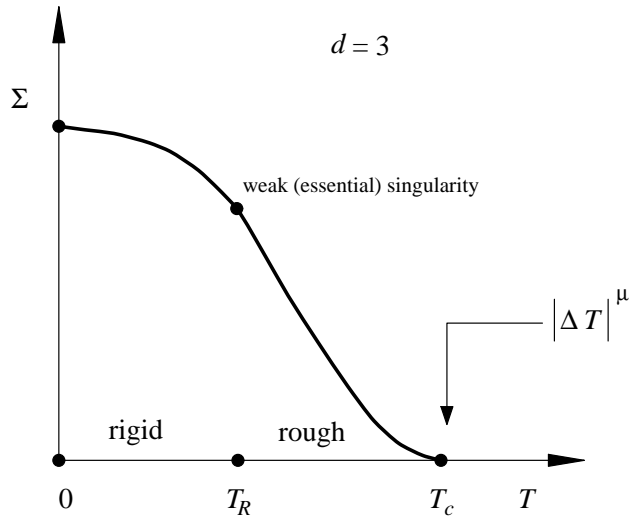

Fig. 6. Surface tension of $d = 3$ systems.

The surface stiffness coefficient in $2d$ is shown schematically in Figure 5. Near $T_c$, where the fluctuations become isotropic, the added second-derivative term in (2.3) is asymptotically smaller than the first term. The surface stiffness simply follows the surface tension. However, as $T \to 0$, the second term dominates. In the expression $kT\kappa$, plotted in Figure 5, the surface-tension contribution is finite as $T \to 0$. However, the second-angular-derivative contribution actually diverges according to $\exp(const/kT)$, as $T \to 0$, where the constant is system-dependent.

The surface tension of the $3d$ systems, see Figure 6, behaves similarly to the $2d$ case. Near $T_c$, (2.4) applies, with typical $\mu$ values in the range $1.0^+$ to $1.5$, depending on the universality class of the critical point at $T_c$, at which the phases loose their distinction and the interfacial structure merges with the rest of the homogeneous system. However, $3d$ interfaces of *lattice (crystalline) systems only*, become rigid for $0 \leq T < T_R$, where the roughening temperature is of order $T_c/2$. The surface tension has an essential singularity at $T = T_R$.

The surface stiffness coefficient in $3d$ is shown schematically in Figure 7. As will be further described later, in Sections 3-5, it is defined only for $T_R \leq T < T_c$. Indeed, it turns out that in the "rigid" regime we have the angular dependence in the form $\sigma(T, |\theta|)$, where $\theta$ is measured with respect to one of the lattice axis, so that (2.1)-(2.3) cannot be used. For a more detailed description of the behavior near $T_R$ the reader is addressed to a review on the roughening transition (Ref. 9) which is believed to belong to the Kosterlitz-Thouless universality class of $d = 2$, $O(2)$-symmetric bulk $XY$ systems.

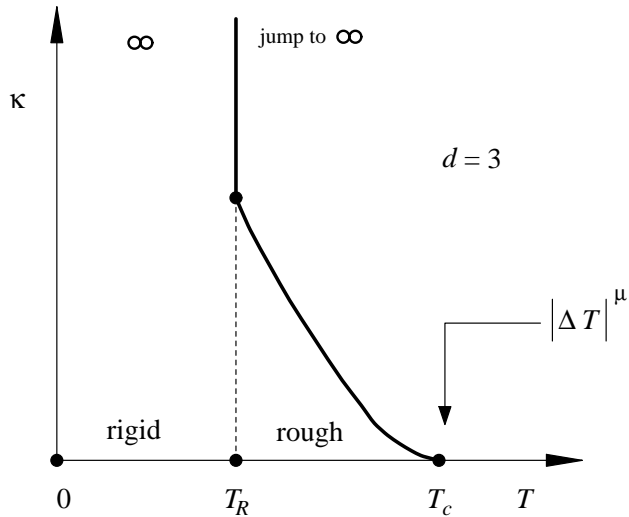

Fig. 7. Surface stiffness coefficient of lattice $d = 3$ systems.

## 3. Capillary Wave Model

In this section we describe the capillary wave model of the longest wave-length, "soft-mode" fluctuations of rough interfaces. Our formulation will be in the $2d$ notation, while extensions to $d > 2$ will be described as appropriate. On large length scales, rough interfaces look sheet-like and can be approximately described by a single valued height function $h(x, \ldots)$, where the unspecified arguments refer to dependence on other transverse coordinates, say $y, z, \ldots$, for $d > 2$.

In the capillary wave model, the energy of the fluctuating interface is approximated by the "macroscopic" energy obtained by intergrating the angle-dependent surface tension over the surface defined by the height function $h$. The $2d$ definitions are illustrated in Figure 8. We note that

$$\theta = \arctan\left(\frac{dh}{dx}\right), \tag{3.1}$$

and so the energy is given by

$$E = kT \int dx \, \sigma[\arctan(dh/dx)] \sqrt{1 + (dh/dx)^2}. \tag{3.2}$$

The square root in (3.2) comes from the element of length,

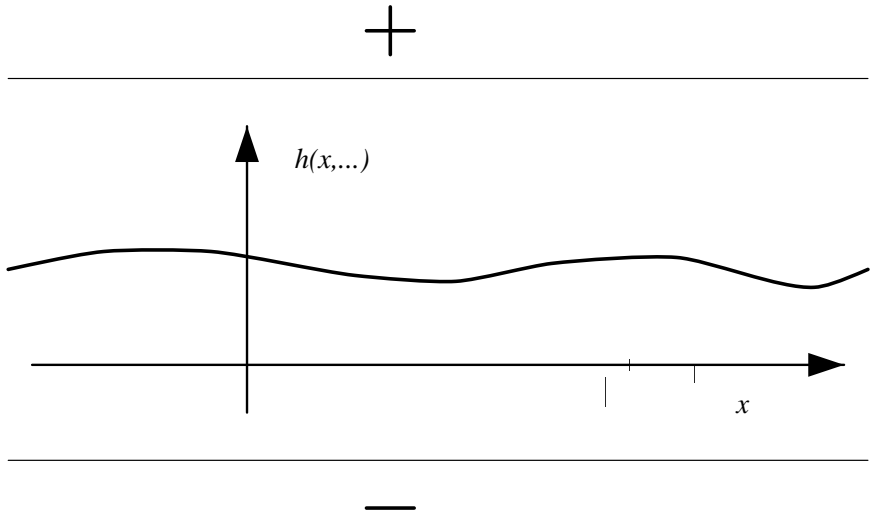

Fig. 8. Capillary wave interface description.

$$d\ell = \sqrt{(dh)^2 + (dx)^2} \, . \tag{3.3}$$

Thus, for $d > 2$ the derivative term under the square root will be replaced by $(\partial h/\partial x)^2 + (\partial h/\partial y)^2 + \ldots$, with $(d-1)$ transverse-coordinate derivatives. The angular dependence in $d > 2$ will also be more complicated, as already mentioned.

The single-sheet description ignoring overhangs, bubbles, and generally assuming that all the "intrinsic" interface structure has been absorbed in the macroscopic surface tension values, is of course approximate. In fact, to make the resulting model tractable, one further assumes that the interface remains largely flat. Thus, the derivatives are small. Specifically, the integrand in (3.2) can be expanded in $(dh/dx)$, keeping the second-order terms,

$$\sigma(0) + \sigma'(0)\left(\frac{dh}{dx}\right) + \frac{1}{2}\kappa\left(\frac{dh}{dx}\right)^2 + \ldots \, . \tag{3.4}$$

The prime here denotes the angular derivative. Note the emergence of the surface stiffness coefficient as the prefactor of the quadratic term, — the origin of the term "surface stiffness."

Integration of the first term yields the constant reference energy proportional to the interface length. The second term only contributes at the end points, the precise specification of which is of no interest. Only the quadratic term depends on the local fluctuating variable $h(x)$. The resulting expression

$$E_{cw}/kT = \frac{1}{2}\kappa \int dx \left(\frac{dh}{dx}\right)^2 , \qquad (3.5)$$

is the standard capillary wave energy, $E_{cw}$, in $2d$.

In higher dimensions, for general interface orientations, expansion to the second order in derivatives with respect to the $(d-1)$ transverse coordinates will produce a mixed-derivative (second order) form due to angular dependence. In fact, a rotation to principal-axes coordinates, $(x, y, z, \ldots) \to (X, Y, Z, \ldots)$, is required to obtain a diagonal form

$$E_{cw}/kT = \frac{1}{2} \int (dX\, dY\, dZ \ldots) \left[ \kappa_X \left(\frac{dh}{dX}\right)^2 + \kappa_Y \left(\frac{dh}{dY}\right)^2 + \ldots \right] . \qquad (3.6)$$

However, in most numerical applications the interfacial orientation is symmetric with respect to the underlying lattice structure. Thus, the original coordinates can be used, and all the stiffness coefficients are equal.

Strictly speaking, the capillary wave energies (Hamiltonians) (3.5), (3.6), must be supplemented by cutoff prescriptions to have finite expressions for thermodynamic average quantities, when these energies are used in Boltzmann factors. Otherwise, these Gaussian models are exactly solvable. Specifically, in $2d$ the capillary fluctuations are indeed strong. Calculations with the Hamiltonian (3.5) suggest that a $2d$ interface "viewed" on the length scale $L$ (for instance, pinned by its ends at separation $L$), fluctuates over vertical ($h$-direction, see Figure 8) span of order $\sqrt{L}$. Detailed Gaussian calculations for all $d$, and references to literature, can be found for instance in Refs. 13, 3.

The case $d = 3$ turns out to be marginal: the interface only spans $\sim \sqrt{\ln L}$ in the vertical direction, when viewed on length scales $L$ (in both transverse directions). For $d > 3$, the capillary span is of order 1. Thus, the model must be applied with great caution. When capillary fluctuations are comparable to other fluctuations in the system, there is no obvious reason to single them out. Some indirect evidence for characteristically Gaussian, capillary fluctuations may come from studies of certain finite-size effects. However, the form of these has been a subject of a recent controversy which will not be reviewed here (but see other articles in these Proceedings).

## 4. Transfer Matrix Method and Capillary Wave Fluctuations

Interfacial parameters, $\sigma$, $\kappa$, etc., can be measured by various methods. As already mentioned the most direct approach is based on comparing a system with an interface with another system where the boundary conditions do not impose any interfacial structure. While finite-size effects are usually corrections to the leading "bulk" results, they can also be used to measure interfacial parameters by numerical techniques.

In this section we describe yet another method of measuring the interfacial properties, based of spectral features of the transfer matrix (Refs. 14-15). No review of the transfer matrix method will be presented here (see Ref. 16-18). Instead, we offer the following general outline. The transfer matrix method is based on summing up over the system configurations by adding consecutive slices of size $b \times A$, in the cylindrical geometry of $(d-1)$-dimensional cross-section of area $A$, and axial length $L_\parallel$. Thus, the partition function is calculated by summing up over configurations of $L_\parallel/b$ slices. In numerical calculations $b$ is usually just one lattice spacing, while the cross-section shape is square or other lattice-aligned shape.

Boltzmann factors incorporating the interaction energy of the added slice with the previous one, as well as its internal energy (to have a symmetric transfer matrix the latter is usually replaced by the mean over two slices), are arranged in a matrix labeled by slice configurations. Calculation of the partition function by summing over system configurations then reduces to matrix multiplication, with summation over each matrix index corresponding to summing over the configurations of one slice of axial length $b$.

Of particular interest are the eigenvalues of this "transfer matrix,"

$$\Lambda_0 > \Lambda_1 > \Lambda_2 > \ldots, \qquad (4.1)$$

some of which can be multiply-degenerate (but not $\Lambda_0$). The reduced (per $kT$) free-energy density of the infinite system, $L_\parallel = \infty$, is given by

$$f = -(\ln \Lambda_0)/(bA) . \qquad (4.2)$$

This free energy of course depends on the cross-section size, shape, boundary conditions, etc., as corrections. It approaches the bulk free-energy density, $f(\infty)$, as $A \to \infty$.

Correlation lengths (proportional to the inverse energy gaps in the Quantum-Hamiltonian nomenclature popular in Particle Physics) in the long-cylinder geometry, $L_\parallel = \infty$, describe the exponential decay of correlations at large distances along the cylinder. They are given by

$$\xi_j = \frac{b}{\ln(\Lambda_0/\Lambda_j)} , \qquad (4.3)$$

with $j = 1, 2, \ldots$.

Correlation lengths associated with the longest-length-scale fluctuations will be the largest, i.e., they will have the lowest indices $j$ in (4.3). Generally, the number of slice configurations increases exponentially with the cross-section size. Numerical diagonalization of transfer matrices is thus a large-scale computational problem. However, theoretically one can in some cases express the largest correlation lengths explicitly in terms of few parameters. This description relies on that fluctuations in the long-cylinder geometry are effectively one-dimensional on large length scales. Thus, an effective, few-parameter parametrization of the transfer matrix can be found (Refs. 19, 15) which reproduces correctly the asymptotic form of the largest correlation lengths. As a result, parameters determining large-scale fluctuations, which for systems phase-separated in the bulk, usually turn out to be various interfacial properties, can be probed directly by transfer-matrix calculations. We describe one such application in this section.

Let us consider the long-strip $2d$ geometry shown in Figure 8, with $\pm$ boundary conditions ensuring that an interface is running along the strip. The height of the strip in the transverse direction will be denoted by $H$, and we assume that it spans the distance from $h = 0$ to $h = H$. (The axes in Figure 8 were shifted for clarity.) In the effective, phenomenological description focusing only on the large-scale interfacial fluctuations which are rough (we assume $0 < T < T_c$), the configuration is specified by the height variable $h$. The effective transfer matrix becomes an operator "labeled" by $h$, in the following sense. The vector space on which it acts is that of functions $\psi(h)$, and the boundary conditions which essentially mean that the interface cannot cross the walls, can be shown to correspond to

$$\psi(0) = \psi(H) = 0 \ . \tag{4.4}$$

The effective transfer operator acts on these functions, and it turns out that the $h(x)$-related fluctuation part of the interfacial energy, singled out in relation (3.5), is represented in this language by the operator

$$\exp\left(\frac{b}{2\kappa} \frac{d^2}{dh^2}\right) \ . \tag{4.5}$$

We will not describe in detail the derivation of this result but instead emphasize its physical meaning and implications. The "free-field Hamiltonian" appearing in the exponent of (4.5) naturally corresponds to the fact that interfacial fluctuations are diffusion-like in the transverse direction, within the capillary wave model, with the diffusion rate inversely proportional to the surface stiffness coefficient $\kappa$.

As in the derivation of (3.5), here there are also contributions which do not arise from the fluctuations included in the capillary wave description. These account for the overall free energy density of the bulk, $kTf(\infty)$, of the interface (flat), $kT\sigma$ (measured per unit length), as well as two $\pm$-wall free-energy terms, equal due to the $\pm$ symmetry. The latter will be denoted by $kTw$ per unit axial length. Thus, the effective transfer operator takes the form

$$\exp\left[-bHf(\infty) - b(\sigma + 2w) + \frac{b}{2\kappa}\frac{d^2}{dh^2}\right] \ . \tag{4.6}$$

The eigenvalues are all nondegenerate,

$$\Lambda_j = \exp\left[-bHf(\infty) - b(\sigma + 2w) - b\frac{\pi^2(j+1)^2}{2\kappa H^2}\right] \ . \tag{4.7}$$

The free-energy relation (4.2) with $A = H$ then gives

$$f(H) = f(\infty) + \frac{\sigma + 2w}{H} + \frac{\pi^2}{2\kappa H^3} + o\left(H^{-3}\right) . \tag{4.8}$$

Thus, we derived the leading-order capillary correction to the free energy in the strip geometry. While the first two terms are the standard reference bulk, surface, and interface free energies, the fluctuation contribution of order $H^{-3}$ is determined by the surface stiffness coefficient. Note that the higher-order terms are not known generally (the lower-case $o$ denotes "less than order of"). They need not be fully capillary wave dominated.

The correlation lengths follow from (4.3) as

$$\xi_j = \frac{2\kappa H^2}{\pi^2 j(j+2)} + o\left(H^2\right) . \tag{4.9}$$

This form of course applies only to the leading correlation lengths, $j = 1, 2, \ldots$; the precise criterion is described elsewhere (Ref. 15). The largest correlation lengths thus provide a direct measure of $\kappa$.

Various other results and observations derivable from the effective transfer-matrix formulation are not reviewed here. Let us only mention that the "ground state" wave function $\psi_0(h)$ is real, non-negative, and provides the probability distribution of finding the interface at $h$. There are other differences between the quantum mechanical time evolution formulation, in "Euclidian time," where the wave functions yield probability amplitudes, and the transfer matrix formulation in Statistical Mechanics. Thus, caution and care must be exercised when methods and results are taken over from one field to another.

In numerical studies, systems with antiperiodic boundary conditions are frequently preferred to those with walls because the former preserve some of the translational-invariance features, and there are no wall energies, $w = 0$. The transfer matrix formulation can be easily extended to antiperiodic boundary conditions. However, it turns out that these boundary conditions effectively induce a topology of $\mathrm{mod}(2H)$ in the strip: interfacial height $h$ can be reached with $+$ above, $-$ below the interface, or with $-$ above and $+$ below. Two rotations of the interface around the system, by vertical displacement, are required to repeat the same vertical-slice configuration of the $\pm$ phases.

For antiperiodic boundary conditions, the relation (4.4) is replaced by

$$\psi(h) = \psi(h + 2H) . \tag{4.10}$$

The results (4.8) and (4.9) are then replaced by

$$f(H) = f(\infty) + \frac{\sigma}{H} + O\left[\exp\left(-H/\xi_{bulk}\right)\right] , \tag{4.11}$$

$$\xi_{j>0} = \frac{2\kappa H^2}{\pi^2 j^2} + o\left(H^2\right) , \tag{4.12}$$

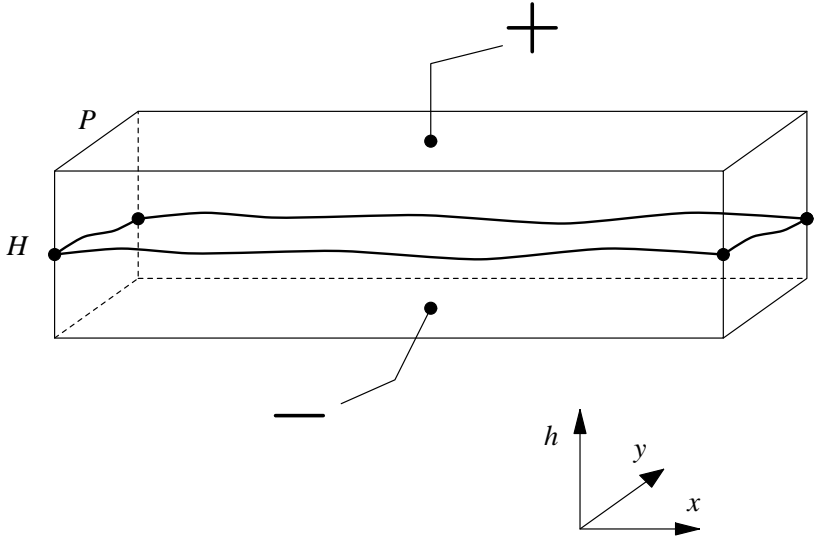

Fig. 9. Interface in a long-cylinder geometry in $3d$.

where the correction in (4.11) is believed to be exponentially small. Other differences include double-degeneracy of $\Lambda_{j>0}$, etc.

Exact calculations for $2d$ Ising-model strips yield expressions for the correlation lengths (Refs. 20-23) which check the capillary wave model predictions and also verify the limits of validity of the capillary approximation which will be briefly discussed at the end of this section.

We now turn to the $3d$ case, in the rough-interface regime, $T_R \leq T < T_c$. The $\pm$ long-cylinder geometry is shown schematically in Figure 9. The system size in the direction perpendicular to the $(h,x)$ plane (marked by $y$ in the figure) will be denoted by $P$. We assume that the boundary conditions along the third, $y$-axis direction are periodic or free, so that the only interface-imposing boundary conditions remain at the top and bottom walls (at $h = 0, H$), as in the $2d$ case. We also assume that $P/H = O(1)$.

The longest wave-length capillary modes then "propagate" along the cylinder. They correspond to fluctuations that leave the interface relatively flat along the $y$ direction; recall that in regions of axial size $\Delta x = O(P)$ the interface is only spread $O\left(\sqrt{\log P}\right)$. However, the interface can fluctuate up and down as a whole, on large $x$-direction length scales which are precisely the largest correlation lengths that we want to estimate. However, these fluctuations are effectively two-dimensional, with the effective stiffness coefficient $P\kappa_x$.

We adopt the notation suggested by (3.6): the $3d$ stiffness coefficient will be denoted by $\kappa_x$. Then the effective value for the $2d$-like fluctuations is $\kappa = P\kappa_x$, and we can use the previously derived leading-order expressions (4.8), (4.9), (4.11), (4.12) with this identification. Note that the units of $\kappa$ are $(\text{length})^{-(d-1)}$ so that the effective $2d$ stiffness coefficient has the correct units.

Recent numerical study (Ref. 24) used the capillary wave model results described here to estimate the surface stiffness coefficient of the $3d$ Ising model in the rough-interface regime. Both the exact $2d$ Ising results mentioned earlier and the numerical $3d$ studies indicate that on approach to the "dangerous" points in the phase diagram, $T = T_c$ and, in $3d$, $T = T_R$, the expressions derived above must be used with caution. Specifically, for $T \to T_c^-$, the forms proposed will "cross-over" to the full finite-size scaling description. The corrections in inverse powers of $H$ in fact become functions of $\xi_{bulk}/H$, and the capillary wave expressions break down as $\xi_{bulk}$ diverges at criticality.

At the roughening temperature, the finite-size scaling forms are not well understood, and in fact it is not even clear if they exist. Thus, we have no theory for the form of cross-over from the rough-interface capillary wave regime to the rigid-interface regime (see the next section), in finite-size systems.

## 5. Rigid Three-Dimensional Interfaces at Low Temperatures

In order to complete our discussion of interfacial fluctuations, we now address fluctuation properties of interfaces in the low-temperature regime of $3d$ lattice models (or crystalline-phase systems) which, as already pointed out are "rigid." More precisely, at low temperatures the interfaces prefer to align with the lattice structure due to energy gain. Thus, an interface which on the average is oriented along one of the underlying lattice axes, will follow this direction not just on the average but also microscopically. Of course, for any nonzero temperature the interface will be "dressed" by thermal fluctuation features such as pits, overhangs, bends, etc. However, these are no soft-mode excitations as for rough interfaces. Rather, these thermal fluctuations are on length scales comparable to the bulk correlation lengths of the coexisting phases.

The two interfacial fluctuation regimes are separated by a sharp phase transition, the roughening transition (Refs. 9), which is in the Kosterlitz-Thouless universality class. However, our discussion will be focused on fluctuations away from $T_R$. In fact, we assume that the temperature is fixed in $0 < T < T_R$, and the system sizes are large enough so the the "critical-like" aspects of the roughening as a phase transition have no effect. (See also note at the end of the preceding section.)

The nature of the rigid-interface fluctuations is best revealed by considering the so-called "vicinal" interfaces which are inclined a small angle $\theta$ with respect to the preferred symmetric lattice direction. Such as interface is schematically shown in the right panel of Figure 10. As mentioned, rigid interfaces tend to follow lattice directions. For simplicity let us assume that the lattice is cubic, of spacing $a$, as shown schematically in the lower right corner of the picture. The interface shown in Figure 10 is forced by boundary conditions to raise $(\Delta h/a)$ lattice spacings when it spans the sample length $L$. Rough interfaces follow their average inclination, see the left panel in Figure 10, accompanied by large, capillary fluctuations on scales exceeding the lattice spacing $a$. However, rigid interfaces actually achieve the appropriate inclination angle by having $(\Delta h/a)$ microscopic steps.

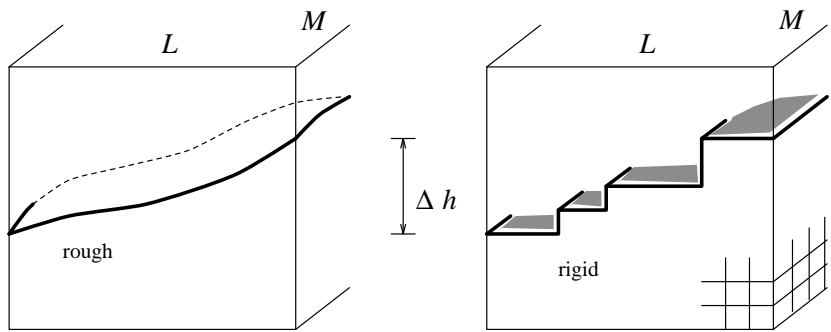

Fig. 10. Rough vs. rigid interfaces in $3d$.

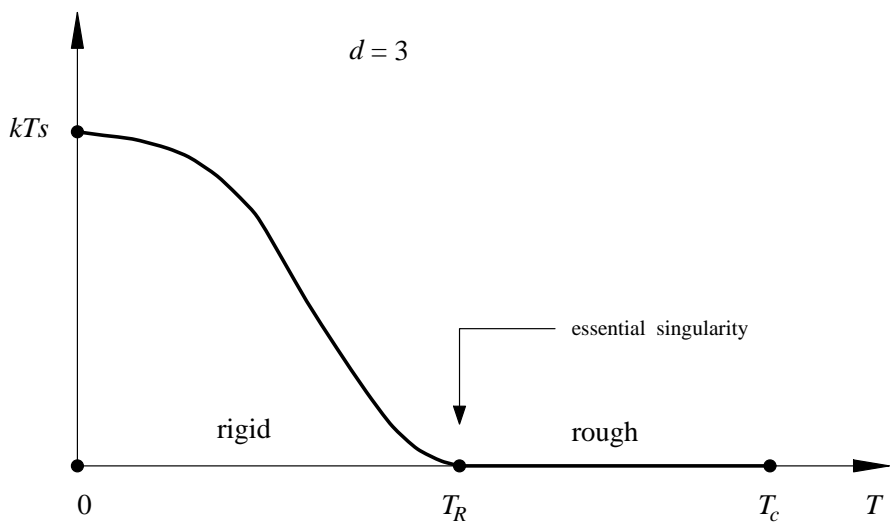

Fig. 11. Step free energy of rigid interfaces in $3d$.

Each step has excess free energy proportional to its length, $M$, where we assume that the sample dimensions in the horizontal cross-section are $L \times M$, see Figure 10. Let $s(T)$ denote this excess free-energy per unit length (of a step) and per $kT$. Actually, $s(T)$ is defined only with respect to symmetric lattice directions, and in fact not only its $T$-dependence may vary for different directions, but the roughening temperature, $T_R$, itself may differ for various lattice orientations as well. The behavior of this *step free energy* is shown schematically in Figure 11. At low temperatures, $kTs$ is a finite quantity. As $T \to T_R^-$, the step free energy vanishes with an essential singularity. Near the roughening transition the length scale $s^{-1}$ diverges and plays the role of a correlation length similar to the correlation length at the Kosterlitz-Thouless transition in $2d$ $XY$ systems on the high-temperature side of that transition (Refs. 9, 25-26). For $T > T_R$, formally $s = 0$.

The total free-energy cost of the rigid interface shown in Figure 10 is given by

$$\Delta F = kT\sigma(0)LM + kTsM(|\Delta h|/a) + \ldots . \qquad (5.1)$$

Since the inclination angle $\theta$ is

$$\theta = \arctan(\Delta h/L) , \qquad (5.2)$$

the angular-dependent surface tension takes the form

$$\sigma(\theta) = \frac{\Delta F}{LM/\cos\theta} = \sigma(0) + sa^{-1}|\theta| + \ldots . \qquad (5.3)$$

Note for comparison that the leading angle-dependent free energy term for rough interfaces in symmetric lattice directions is quadratic in $\theta$, as suggested by (2.1) with $\sigma'(0) = 0$. However, in the rigid-interface regime the $\theta$-dependence is no longer analytic, and the leading, order $|\theta|$ free-energy correction is determined by the step free energy and is explicitly lattice-spacing dependent.

Of cause the above arguments only apply for very small angles, when the steps are well-separated. The correction terms shown by ... in (5.1) and (5.3) will be contributed by the interplay of two effects. Firstly, at any nonzero temperature the steps, as well as the flat interface portions, see Figure 10, will be "dressed" by thermal-fluctuation features. Those on scales of the correlation lengths of the two coexisting phases can be considered absorbed in the definition of $s(T)$. However, the steps when viewed "from above" in Figure 10, will not be just straight lines of length $M$. They can have kink-like shifts along the $L$-direction and will therefore behave random-walk like. Even though "kinks" cost energy, and as a result the effective two-dimensional stiffness of these soft-mode-like step fluctuations will be large, it will definitely not be infinite.

Secondly, for finite angles $\theta$, the step wandering just described will contribute a long-range term (Refs. 9) in *step-step interactions*. Short-range step-step interactions also arise from energy and entropy changes associated with step contacts. These interactions determine the corrections in (5.1), (5.3). Step interaction and fluctuation phenomena are accounted for semiquantitatively within the *terrace-ledge-kink* model of rigid-interface fluctuations. The reader should consult the more specialized review (Ref. 9) for details, further literature citations, and for a description of how the large-angle variation of the interfacial free energy actually determines crystal shapes.

Direct estimates of the step free energy have been reported in Monte Carlo studies of $3d$ Ising models with $\Delta h = a$ (Refs. 25-26). In the rest of this section we describe how the transfer matrix spectrum is changed in the rigid-interface regime. We find that the spectral gaps (inverse correlation lengths) are determined by $s(T)$ and thus provide a direct measure of this interfacial property.

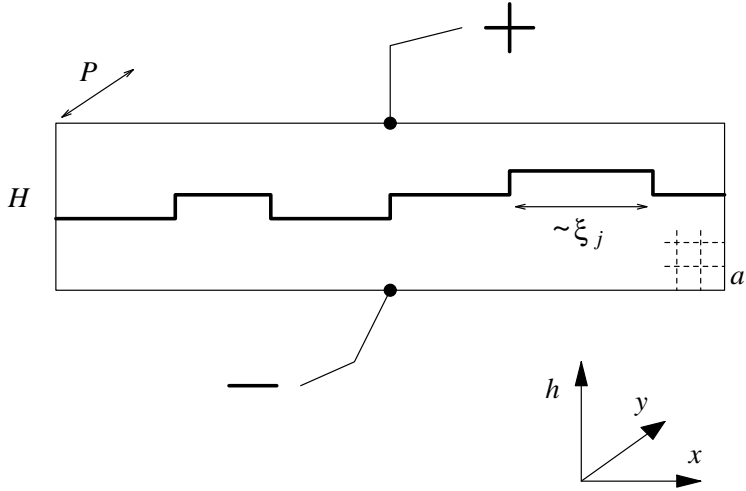

Fig. 12. Directed-walk model of rigid interface fluctuations in $3d$ cylinders.

In the long-cylinder geometry, of cross-section size $H \times P$, see Figure 9, we now consider the regime of $0 < T < T_R$. The interface will no longer be rough as in Figure 9. Instead, its will be microscopically flat over large length scales, measured by the leading correlation lengths $\xi_j$. The dominant fluctuations will be well-separated steps (we will see later that $\xi_j \gg H, P$). Such fluctuations can be modeled by a solid-on-solid model, of steps (forward, up and down) forming a directed random walk, as illustrated in Figure 12. Such models have been used extensively for two-dimensional interfaces (Refs. 3, 5, 11).

However, as a model of the "cross-sectional" form of the rigid $3d$ interface the directed walk has a simplifying feature. Indeed, the transverse, up and down, steps of length $a$ are extremely improbable. Their free-energy cost is represented by the exponentially small (in the system size $P$) Boltzmann factor

$$\exp\left(-Ps + \ldots\right). \tag{5.4}$$

The Boltzmann factor per each horizontal step of length $b$, — the slice size, — is given by the expression that incorporates the nonfluctuation, "reference" free-energy contributions,

$$\exp\left[-bHPf(\infty) - bP\sigma + \ldots\right]. \tag{5.5}$$

The correction terms in (5.4) and (5.5), marked $\ldots$, depend on finite-size effects associated with the boundary conditions in the $H$ and $P$ directions; they will not be described here (see Ref. 15).

Note that the weight of the transverse steps is small as compared to the "no step" Boltzmann factor, 1. Longitudinal weights are also numerically small, but for directed walks their number (per unit system length) is not a fluctuating quantity. Due to small weights (5.4), one can show that the results do not depend on the details of the walk model chosen. For instance, they are the same if only single or any number of transverse steps are allowed at each fixed $x$-coordinate value (these are known as, respectively, restricted and unrestricted $2d$ solid-on-solid models).

The *effective transfer matrix* acts now in the space determined by the allowed $h$-coordinate values which are no longer continuous as for rough interfaces. Indeed, as illustrated in Figure 12, the allowed $h$ values will be discretized in steps of $a$. Thus, the interface can have $O(H/a)$ height values. The effective transfer matrix acts on vectors with this number of elements and therefore has $O(H/a)$ eigenvalues. Explicit calculations are rather technical, and we only summarize here the resulting expressions for the correlation lengths.

The eigenvalues of the effective transfer matrix form a distinct multiplet associated with the rigid-interface fluctuations. Thus, the leading $O(H/a)$ eigenvalues are in fact split in a very narrow band, $\sim \exp(-Ps)$, and they are well separated (by spectral gaps of order 1) from the rest of the spectrum (of the full transfer matrix). For $\pm$ boundary conditions, shown in Figure 12, we find the expression,

$$\xi_{j>0} = b\gamma_{\pm}(T)P^w e^{Ps(T)} \left[\sin\frac{\pi a(j+2)}{2(H+a)} \sin\frac{\pi a j}{2(H+a)}\right]^{-1}, \qquad (5.6)$$

where the eigenvalues are labeled by $j = 0, 1, \ldots, (H/a) - 1$. Thus, only the $(H/a)$ largest, nondegenerate eigenvalues are determined by $s(T)$ via the leading, exponential size-dependence of the (inverse) gaps.

The prefactor quantities, $\gamma_{...}(T)$ and the power $w$, are boundary condition and shape-dependent. The power $w$ has been related to finite-size corrections due to the finite extent $P$. For, respectively, periodic and free boundary conditions in the $P$ direction, the predicted values are $w = \frac{1}{2}$ and 0. However, these results are essentially just conjectures.

For antiperiodic ($ap$) boundary conditions, the results are

$$\xi_{j>0} = b\gamma_{ap}(T)P^w e^{Ps(T)} \left(\sin\frac{\pi a j}{2H}\right)^{-2}, \qquad (5.7)$$

where now the index covers the range $j = 0, 1, \ldots, H/a$, and furthermore all but the $j = 0, H/a$ eigenvalues are doubly-degenerate. The power $w$ is the same as in (5.6). No detailed predictions are available for the functions $\gamma_{...}(T)$.

Numerical studies of the $3d$ Ising model below $T_R$ (Ref. 24) checked the leading-order step-free-energy dependence, $\sim \exp(Ps)$. The expressions surveyed here work well except on approach to the roughening temperature $T_R$. In this limit the sizes needed to observe the asymptotic regime of validity of the "microscopic" expressions (5.6), (5.7), etc., become large and cannot be reached in numerical simulations. As emphasized in Section 4, there is presently no cross-over theory to describe the behavior near $T_R$.

**Acknowledgements**

The author has enjoyed collaboration with Professor Nenad M. Švrakić on topics in interfacial fluctuations and related fields. He wishes to thank the conference organizers for a pleasant and productive meeting.

This work was partially supported by the Science and Engineering Research Council (UK) under grant number GR/G02741. The author also wishes to acknowledge the award of a Guest Research Fellowship at Oxford from the Royal Society.